\documentclass[12pt,preprint]{aastex}

\received{}
\revised{}
\accepted{}

\shorttitle{Orientation of Pairs of Spiral Galaxies}
\shortauthors{Buxton \& Ryden}

\begin{document}
\title{Relative Orientation of Pairs of Spiral Galaxies in
the Sloan Digital Sky Survey}
\author{Jesse Buxton\altaffilmark{1} \& Barbara S. Ryden\altaffilmark{2}}
\affil{Department of Astronomy, The Ohio State University,
Columbus, OH 43210}
\altaffiltext{1}{Department of Physics, The Ohio State University,
Columbus, OH 43210}
\altaffiltext{2}{Center for Cosmology \& Astro-Particle Physics,
The Ohio State University, Columbus, OH 43210}
\email{buxton.45@osu.edu, ryden@astronomy.ohio-state.edu}

\begin{abstract}

We find, from our study of binary spiral galaxies in the
Sloan Digital Sky Survey Data Release 6, that the relative
orientation of disks in binary spiral galaxies is consistent with
their being drawn from a random distribution of orientations.
For 747 isolated pairs of luminous disk galaxies, the
distribution of $\phi$, the angle between the major axes
of the galaxy images, is consistent with a uniform distribution
on the interval $[0^\circ , 90^\circ ]$. With the assumption
that the disk galaxies are oblate spheroids, we can compute
$\cos\beta$, where $\beta$ is the angle between the rotation
axes of the disks. In the case that one galaxy in the
binary is face-on or edge-on, the tilt ambiguity is resolved,
and $\cos\beta$ can be computed unambiguously. For 94 isolated
pairs with at least one face-on member, and for 171 isolated
pairs with at least one edge-on member, the distribution of
$\cos\beta$ is statistically consistent with the distribution
of $\cos i$ for isolated disk galaxies. This result is consistent
with random orientations of the disks within pairs.

\end{abstract}


\section{INTRODUCTION}
\label{sec-intro}

The structure of a spiral galaxy is closely tied to its angular
momentum distribution. Thus, understanding galaxy formation requires
knowledge of how galaxies acquire their angular momentum.
The tidal torque model for the acquisition of angular momentum
by galaxies \citep{ho49,pe69} states that protogalaxies acquire
their angular momentum through the tidal interaction between
the protogalaxy and the surrounding nonuniform matter distribution.
Spiral galaxies are particularly useful for testing how galaxies acquire
angular momentum. Spiral galaxies contain relatively cold, thin disks of
stars, gas, and dust, supported primarily by rotation rather than velocity
dispersion. These thin disks are unlikely to have suffered a major
merger (or large numbers of minor mergers) since their formation;
thus, the orientation of a disk's spin angular momentum is unlikely
to have changed drastically since the material that formed it
was initially torqued.

Binary spiral galaxies are particularly useful test cases for
the acquisition of angular momentum. A ``binary spiral galaxy''
can be defined as a pair of spiral galaxies that are relatively
close to each other, but are much farther away from other
galaxies of comparable or greater mass. Our galaxy and M31,
using this general definition, can be thought of as a binary
spiral galaxy. They are separated by $\sim 0.8 {\rm\,Mpc}$,
and M33, the next brightest galaxy in the Local Group, is
$\sim 2$ magnitudes fainter than our galaxy or M31. If
a pair of spiral galaxies is a totally isolated
system, with zero net angular momentum initially, then
conservation of angular momentum states that the net
angular momentum remains zero after the two galaxies
have spun each other up. If the net angular momentum
of the binary system is, in fact, zero, then the
relative orientation of the internal angular momentum
vectors of the two galaxies can range from anti-parallel,
if the two galaxies are on radial orbits relative to each
other, to parallel, if the orbital angular momentum is
large.

In the real universe, binary spiral galaxies are
never completely isolated. \citet{go78}, for instance,
in their study of the angular momentum of the Local Group,
conclude that most of the angular momentum of our galaxy
was produced by M31, but that the tidal torquing by nearby
groups, such as the Maffei group ($d \sim 3 {\rm\,Mpc}$) and
M81 group ($d \approx 3.6 {\rm\,Mpc}$), was not negligible.
Thus, we might expect the distribution of the angle $\alpha$ between
the internal angular momentum vectors of the two galaxies
in a pair to depend on the number of massive galaxies within
several megaparsecs of the pair. In addition, the distribution
of $\alpha$ will also depend on the evolutionary history of
binary spiral galaxies since they initially attained their
angular momentum by tidal torquing. For instance, the
merger timescale for a bound pair of galaxies depends
on the ellipticity $\varepsilon$ of their mutual
orbit \citep{la93,ji08}, with more nearly radial
orbits having a shorter merging timescale. Thus, binary
spiral galaxies with radial orbits will more rapidly merge
to form an elliptical galaxy. In the case that the binary
spiral galaxies are initially isolated, this will result
in preferential destruction of the binaries whose internal
angular momenta are anti-parallel.

Computer simulations have provided additional insight into
the alignment of galaxy spins. In $n$-body $\Lambda$
cold dark matter simulations, for instance, massive
dark halos ($M \gtrsim 3 \times 10^{12} M_\odot$)
have spins that are preferentially oriented in a
direction perpendicular to the mass distribution \citep{pa08},
and have projected major axes that are aligned with
the surrounding galaxy distribution \citep{pz11}; these
alignments persist on scales up to $\sim 30 {\rm\,Mpc}$.
It should be kept in mind, though, that a dark halo is
not necessarily a simple structure; \citet{sc11} show
that in the Millennium-2 $n$-body simulation $\sim 25\%$
of all halos have nearly perpendicular major axes at
large and small radii. \citet{ha10}, in a hydrodynamic
adaptive mesh refinement simulation that permits disks
to form by dissipation, find that at $z = 0$, the spin
of stellar and gaseous disks are well aligned with the
spin of the inner dark halo; however, the alignment
between disk spin and the spin of the entire host
halo is significantly weaker. 

Many studies have been done of the relative orientation
of the spin vectors of neighboring spiral galaxies. Three
different methods, using increasing amounts of information
for each binary spiral galaxy, may be distinguished.

The first method for studying the relative orientation
of disk galaxies uses only the position angle of the
apparent major axis of each galaxy. If a galaxy is
approximated as a rotationally flattened oblate spheroid,
then its apparent major axis is at right angles to the
projection of its spin axis onto the plane of the sky.
The angle $\phi$ between the apparent major axes of the
two galaxies in a binary spiral galaxy, defined so that
$0^\circ \leq \phi \leq 90^\circ$, is thus the angle
between the projections of the spin axes. If the
spin axes of the two spiral galaxies tend to be parallel
(implying either parallel or anti-parallel spin
vectors), then the distribution of $\phi$
will be weighted toward $\phi = 0^\circ$; if the
spin axes tend to be perpendicular, then the
distribution of $\phi$ will be weighted toward
$\phi = 90^\circ$ \citep{so92}. Employing this method,
\citet{sh79}, with a sample of 57 pairs of spirals,
and \citet{ce10}, with 218 pairs of spirals, found
distributions for $\phi$ that were statistically
consistent with being uniform from $0^\circ$ to
$90^\circ$; these results are consistent with, but do not
demand, uncorrelated spin vectors. By contrast, \citet{so92},
using a sample of 390 pairs of spirals, found a significant
excess of pairs with $\phi \sim 90^\circ$, indicating
a tendency for spin axes to be perpendicular.

The second method for studying the relative orientation
of spiral galaxies uses the apparent axis ratio $q$ of
each galaxy image in addition to the angle $\phi$ between
their apparent major axes. If each galaxy is approximated
as an oblate spheroid of intrinsic short-to-long axis ratio
$\gamma$, then the inclination $i$ of a galaxy with apparent
axis ratio $q$ is given by the usual relation
\begin{equation}
\cos i = \left( {q^2 - \gamma^2 \over 1 - \gamma^2} \right)^{1/2} \ .
\label{eq:cosi}
\end{equation}
Now consider a pair of disk galaxies, with inclinations $i_1$ and $i_2$,
that have an angle $\phi$ between their apparent major axes. If the
angular separation of the galaxies is small, then the angle $\beta$
between their spin axes, defined so that $0 \leq \beta \leq 90^\circ$,
is given by the relation
\begin{equation}
\cos \beta = | \sin i_1 \sin i_2 \cos \phi \pm \cos i_1 \cos i_2 | \ .
\label{eq:cosbeta}
\end{equation}
The two-fold ambiguity in $\cos \beta$ is due to the tilt ambiguity
of the individual galaxies. (If the apparent major axis of a disk
galaxy lies in the east-west direction, for example, we don't generally
know whether the northern or the southern half of the disk is closer to us.)
Using this method, \citet{fl93}, with a sample of 586 pairs of galaxies, found
that their spin axes tend to be parallel. However, this method is intrinsically
unable to distinguish between parallel or anti-parallel spin vectors.

The third method for studying relative orientation of spiral
galaxies uses additional information to resolve the tilt
ambiguity and spin ambiguity of disk galaxies. The tilt ambiguity
(``Is the northern or southern half of the disk closer to us?'') can be
resolved by looking at high-resolution images of a disk galaxy and assuming
that dust lanes lie on the outer edge of spiral arms. The spin ambiguity
(``Is the disk rotating clockwise or counterclockwise from our point of view?'')
can be resolved, if the galaxy is close to face-on, by looking
at the chirality of the disk's spiral arms and assuming that
spiral arms are trailing. Either of these two pieces of information --
location of dust lanes and chirality of spiral arms -- can be
replaced by spectroscopic information about which half of the
disk is redshifted relative to the galaxy's nucleus and which
is blueshifted.
The additional information lets us find the angle $\alpha$
between the spin vectors of the galaxies, defined so that
$0 \leq \alpha \leq 180^\circ$. Using this method,
\citet{he84}, with a sample of 31 pairs of galaxies, found that
spin vectors tend to be anti-parallel ($\alpha \sim 180^\circ$).
\citet{oo93}, using a sample of 40 pairs, found that their spin vectors
were uncorrelated. \citet{pe04} used a sample of 46 pairs of galaxies,
largely drawn from the study of \citet{oo93}, but discarded the
information that resolved the spin ambiguity; they then found
a tendency for spin axes to be either parallel or perpendicular,
with a shortage of pairs at $\beta \sim 45^\circ$.

All three methods outlined above for determining relative
orientations have shortcomings. Determining the angle $\alpha$
between the spin vectors of two spiral galaxies requires
high-resolution imaging or spatially resolved spectroscopy.
In practice, this has limited sample size; \citet{he84},
\citet{oo93}, and \citet{pe04} all had $N < 50$ pairs in their
samples. In addition, the assumptions must be made that dust
lanes are on the outside of spiral arms and that spiral arms
are trailing; these assumptions are not invariably true.
Determining the angle $\beta$ between the spin axes requires
knowing only the apparent axis ratio $q$ for each disk and the
angle $\phi$ between their major axes; this can be found from
lower resolution images, and permits larger sample sizes.
However, because of the tilt ambiguity of each disk image, the
distribution of $\beta$ is not, in the general case, determined
uniquely. In addition, the assumptions must be made that disks
are axisymmetric and that their intrinsic short-to-long axis
ratio is known; these assumptions are not perfectly true.
Determining the distribution of $\phi$, the angle between the
apparent major axes of binary spiral galaxies, can be done
for even low-resolution images; smearing by the point spread
function (PSF) will increase $q$ for a galaxy, but will not
strongly affect the position angle unless the PSF is severely
asymmetric. This permits large sample sizes. However, even a
perfect knowledge of the distribution $F(\phi)$ does not
yield a unique solution for $f (\alpha)$, the underlying
distribution of angles between the spin vectors.

The problem of determining the relative orientation of
disks in binary spiral galaxies is a difficult one.
Different investigators, even those using similar techniques
or overlapping data sets, find very different conclusions.
These range from random orientation of spin axes \citep{sh79,oo93,ce10},
to a tendency for axes to be parallel \citep{he84,fl93}, to a 
tendency for axes to be perpendicular \citep{so92}, to a tendency
for axes to be either parallel or perpendicular \citep{pe04}.
Our plunge into these troubled waters begins with selecting
a sample of close pairs of disk galaxies from the Sloan Digital
Sky Survey, as described in Section~\ref{sec-data}. Our
initial analysis, described in Section~\ref{sec-phi}
looks at the distribution of $\phi$, the angle between
apparent major axes, for a relatively large ($N = 747$) sample
of isolated pairs of galaxies. Then, in Section~\ref{sec-beta},
we add information about the apparent axis ratios $q$ to find
the distribution of $\beta$, the angle between rotation axes.
To eliminate the ambiguity in equation~(\ref{eq:cosbeta}), we
look only at the $N = 251$ isolated pairs for which at least
one of the galaxies is nearly face-on ($q > 0.9$) or nearly
edge-on ($q \leq 0.3$). Finally, in section~\ref{sec-disc}, we discuss
the implication of our results for the acquisition and evolution
of angular momentum in binary spiral galaxies.

\section{DATA SELECTION}
\label{sec-data}

We draw our sample of galaxies from the Sloan Digital Sky Survey
Data Release 6 (SDSS DR6), which includes 9853 square degrees of
photometric coverage and 7425 square degrees of spectroscopic
coverage \citep{ad08}. The photometric imaging is done in 5
bands ($ugriz$); in this paper, we will use the imaging data
in the $r$ band. To build our sample, we start by selecting
primary photometric objects that are classified as ``galaxies''
(${\rm obj\_type} = 3$) in the $r$ band and that have spectroscopic data
with spectroscopic confidence parameter ${\rm zConf} > 0.35$. We require
that the redshift of each galaxy lie in the range $0.004 < z < 0.1$; the
lower limit on $z$ eliminates contaminating foreground objects, and
the upper limit reduces the possibility of weak lensing distortions
of apparent shapes, and eliminates in practice the necessity of
applying K-corrections. To ensure that the galaxies in this sample
are well resolved, we include only those images with $\tau > 6.25
\tau_{\rm psf}$, where $\tau$ is the adaptive second-order moment
of the galaxy image and $\tau_{\rm psf}$ is the adaptive second-order
moment of the PSF at the galaxy's location. The
sample of galaxies satisfying these constraints contains
$N = 166{,}853$ individual galaxies.

To extract a subsample of rotationally supported disk galaxies
from our sample of $166{,}853$ galaxies, we start by using the
SDSS parameter ``fracDeV'', which provides a measure of the
central concentration of a galaxy's light profile. For each
galaxy image, two models are fitted. One model has a de
Vaucouleurs profile \citep{dV48}:
\begin{equation}
I(R) = I_e \exp \left( - 7.67 \left[ (R/R_e)^{1/4} - 1 \right] \right) \ ,
\end{equation}
truncated beyond $7 R_e$ to go smoothly to zero at $8 R_e$. The
other model has an exponential profile:
\begin{equation}
I(R) = I_e \exp \left( - 1.68 \left[ R/R_e -1 \right] \right) \ ,
\end{equation}
truncated beyond $3 R_e$ to go smoothly to zero at $4 R_e$.
The SDSS data pipeline then takes the best-fitting de Vaucouleurs
model and the best-fitting exponential model, and finds
the linear combination of the two that best fits the galaxy
image. The fraction of the total flux contributed by the
de Vaucouleurs component is the parameter fracDeV, which
is constrained to lie in the interval $0 \leq {\rm fracDeV}
\leq 1$.

Previous studies using SDSS galaxies have used different
cuts in fracDeV to separate late-type (low fracDeV) galaxies
from early-type (high fracDeV) galaxies. \citet{ch06} and
\citet{sh07}, in their studies of late-type galaxies, found
it useful to choose a sample with ${\rm fracDeV} < 0.5$.
In this paper, however, our aim is to produce a sample
of strongly disk-dominated spiral galaxies. To this end,
we make a more stringent cut in the fracDeV parameter,
requiring ${\rm fracDeV} \leq 0.1$. \citet{un08} demonstrated
that relatively luminous ($M_r \lesssim -19$) galaxies with
${\rm fracDeV} < 0.1$ have the properties expected of dusty,
disk-dominated spiral galaxies: they lie primarily in the
blue cloud of the color--magnitude diagram and they
have colors and fluxes that are dependent on the apparent
axis ratio $q$, with smaller values of $q$ corresponding
to redder, fainter galaxies. Moreover, visual inspection of
SDSS ``postage stamp'' images reveals that galaxies with
$M_r \lesssim -19$ and ${\rm fracDeV} < 0.1$ appear to be
primarily of Hubble type Sbc or Sc \citep{un08}. 

The inclination-dependent dimming found by \citet{un08} in
their sample of ${\rm fracDeV} < 0.1$ galaxies is well fitted
by the relation $\Delta M_r = 1.27 ( \log q )^2$. Because
the dimming by dust is inclination dependent, a simple
flux-limited survey, such as the SDSS spectroscopic survey,
will be biased against disks that are nearly edge-on (low $q$).
To provide a correction for this inclination-related bias,
we require that all galaxies in our final sample would be
above the flux limit if they were seen edge-on. If an
SDSS galaxy with ${\rm fracDeV} < 0.1$ has $q \leq 0.2$,
we assume that it is already edge-on. If it has $q > 0.2$,
we compute its edge-on flux to be
\begin{equation}
m_r ({\rm edge-on}) = m_r ({\rm obs}) + 1.27 [ (\log 0.2)^2 - (\log q)^2 ] \ .
\end{equation}
To enter our corrected flux-limited sample, a galaxy must have
$m_r ({\rm edge-on}) \leq 17.77$, corresponding to the completeness
limit for the SDSS spectroscopic survey for galaxies.

Finally, although disk-dominated spirals are well fitted by
exponential profiles, so are other subclasses of galaxies, such
as dwarf ellipticals.  \citet{un08} found that ${\rm fracDeV} < 0.1$
galaxies dimmer than $M_r \sim -19$ have a $u-r$ color that is
independent of $q$. The lack of inclination-dependent reddening
for the relatively low-luminosity galaxies reflects the fact that
they are blue ($u-r \sim 1.5$) dwarf galaxies in which the stars
and dust are not in orderly thin disks. To eliminate the dwarf galaxies
from our sample, we start by following \citet{un08} in computing an
approximate face-on absolute magnitude for each galaxy:
\begin{equation}
M_r ({\rm face-on}) = M_r - 1.27 ( \log q )^2 \ .
\end{equation}
To exclude the dwarfs, we require that $M_r ({\rm face-on})
\leq -19.4$.
Throughout this paper, when computing distances and
absolute magnitudes, we assume a consensus cosmology
with $H_0 = 70 {\rm\,km}{\rm\,s}^{-1}{\rm\,Mpc}^{-1}$,
$\Omega_{m,0} = 0.3$ and $\Omega_{\Lambda,0} = 0.7$.

Applying the additional criteria that ${\rm fracDeV} \leq 0.1$,
$m_r ({\rm edge-on}) \leq 17.77$, and $M_r ({\rm face-on})
\leq -19.4$, we reduce our sample size from $N = 166{,}853$
galaxies to $N = 32{,}358$ luminous, disk-dominated
spiral galaxies. To select binary spiral galaxies from this
sample, we require that the redshift difference between two
disk galaxies corresponds to a radial velocity difference
$\Delta v \leq 300 {\rm\,km}{\rm\,s}^{-1}$, and that the
projected separation between them be $R_p \leq 1 {\rm\,Mpc}$.
To ensure that the binary system is relatively isolated, we
also require that no other galaxy in the SDSS DR6 spectroscopic
survey be within $\Delta v = 300 {\rm\,km}{\rm\,s}^{-1}$
and $R_p = 1 {\rm\,Mpc}$ of either spiral galaxy in the
binary system. These criteria leave us with $N_p = 747$
isolated binary spiral galaxies. If we had chosen
$\Delta v = 600 {\rm\,km}{\rm\,s}^{-1}$ as our critical
radial velocity difference (both for defining binaries
and for defining isolated binaries), then we would have
ended with $N_p = 699$ isolated binary spiral systems, and
the qualitative results of our study would have been
unchanged.

As a control, it will be useful to have a sample
of luminous disk galaxies that are not part of a binary
spiral system, and that are relatively isolated from bright
galaxies of all types. To select a sample of isolated
disks from our sample of $N = 32{,}358$ luminous,
disk-dominated spiral galaxies, we require that no
other galaxy in the SDSS DR6 spectroscopic survey be
within $\Delta v = 300 {\rm\,km}{\rm\,s}^{-1}$ and
$R_p = 1 {\rm\,Mpc}$ of the spiral galaxy in question.
These criteria leave us with $N_i = 16{,}814$ isolated
spiral galaxies. However, it is known that Sloan galaxies
in close pairs have a significantly different distribution
in redshift and stellar mass than non-paired Sloan
galaxies \citep{el08,pe09}. Thus, to ensure that the
control galaxies are similar in their properties to
the galaxies in pairs, we adapt the method of \citet{pa11}
for finding a \emph{matched} control sample. We first
go down the list of $2 N_p = 1494$ disk galaxies in
pairs, and find the isolated galaxy which best matches
the redshift of each paired galaxy; once an isolated
galaxy is matched to a paired galaxy, it is removed
from further consideration. This gives a sample of
1494 isolated galaxies that are very closely matched
in redshift to the galaxies in pairs. To increase
the number of galaxies in our matched control sample,
we iterate this procedure until the redshift distribution
of the matched control sample deviates significantly
from that of the galaxies in pairs. We found that 7
iterations gave a KS probability of $P_{\rm KS} = 0.998$
for comparison of the redshift distributions; going
to an 8th iteration dropped the probability to
$P_{\rm KS} \approx 0.23$. Thus, we adopt a matched
control sample with $N_c = 7 \times 1494 = 10{,}458$
galaxies.

\begin{figure}
\plotone{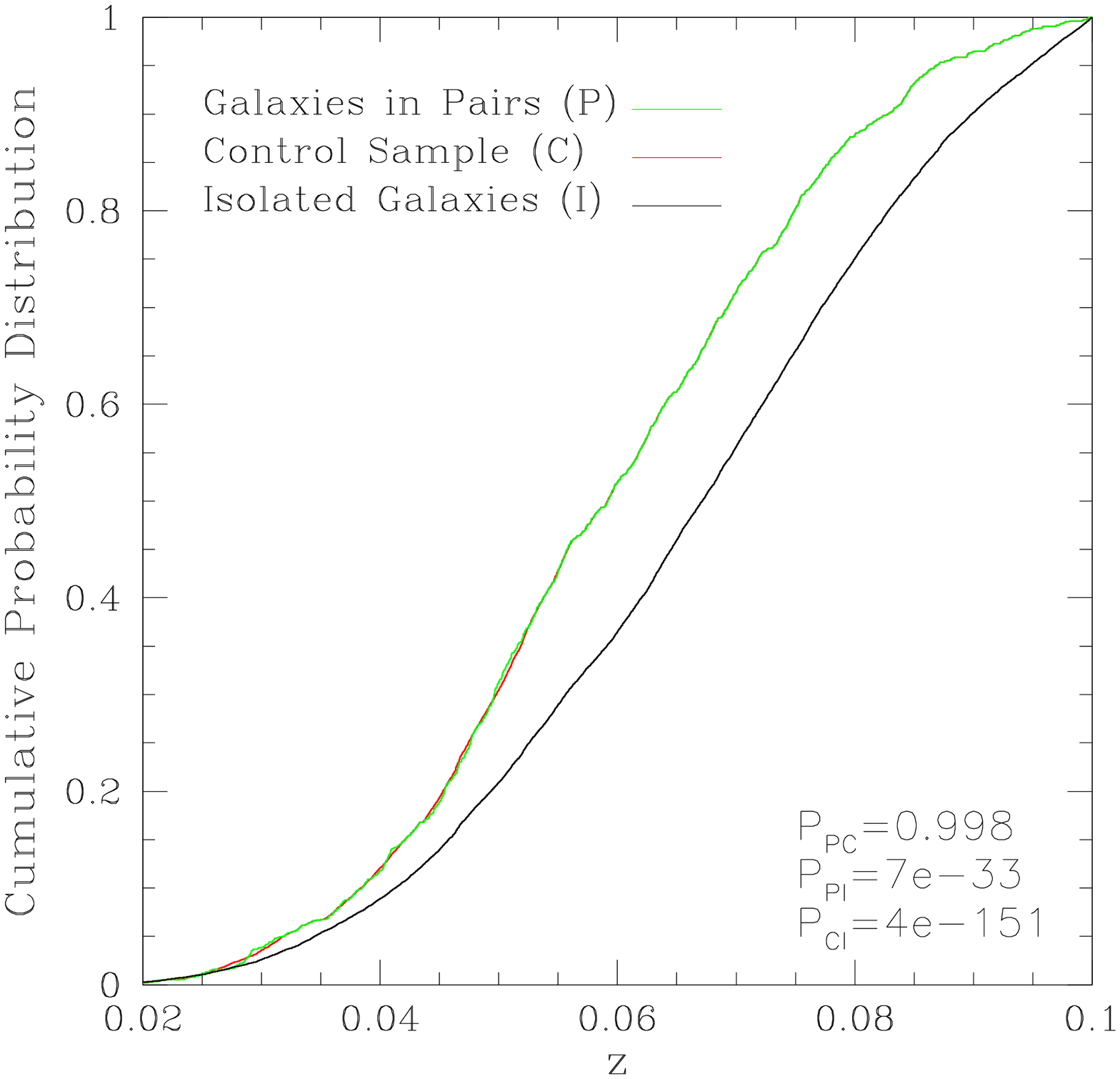}
\caption{The cumulative distribution of redshift $z$ for
the $1494$ disk galaxies in pairs (green), for the
matched control sample of $7 \times 1494 = 10{,}458$
isolated disk galaxies (red), and for the complete
sample of $16{,}814$ isolated disks (black).
}
\label{fig:1}
\end{figure}

Figure~\ref{fig:1} shows the cumulative distribution of
redshifts for the disk galaxies in pairs (green line),
for the matched control sample (red line), and for the
complete unmatched sample of $16{,}814$ isolated galaxies
(black line). In this plot, the redshift distributions
for the disks in pairs and the matched control sample
are nearly indistinguishable (and in fact yield a KS
probability of nearly one). Note that although we
selected our matched control sample on the basis
of redshift, and not on both redshift and estimated
stellar mass, as \citet{pa11} did, we find that selecting
purely on redshift gives similar (but not identical)
distributions in color and absolute magnitude.

\begin{figure}
\plottwo{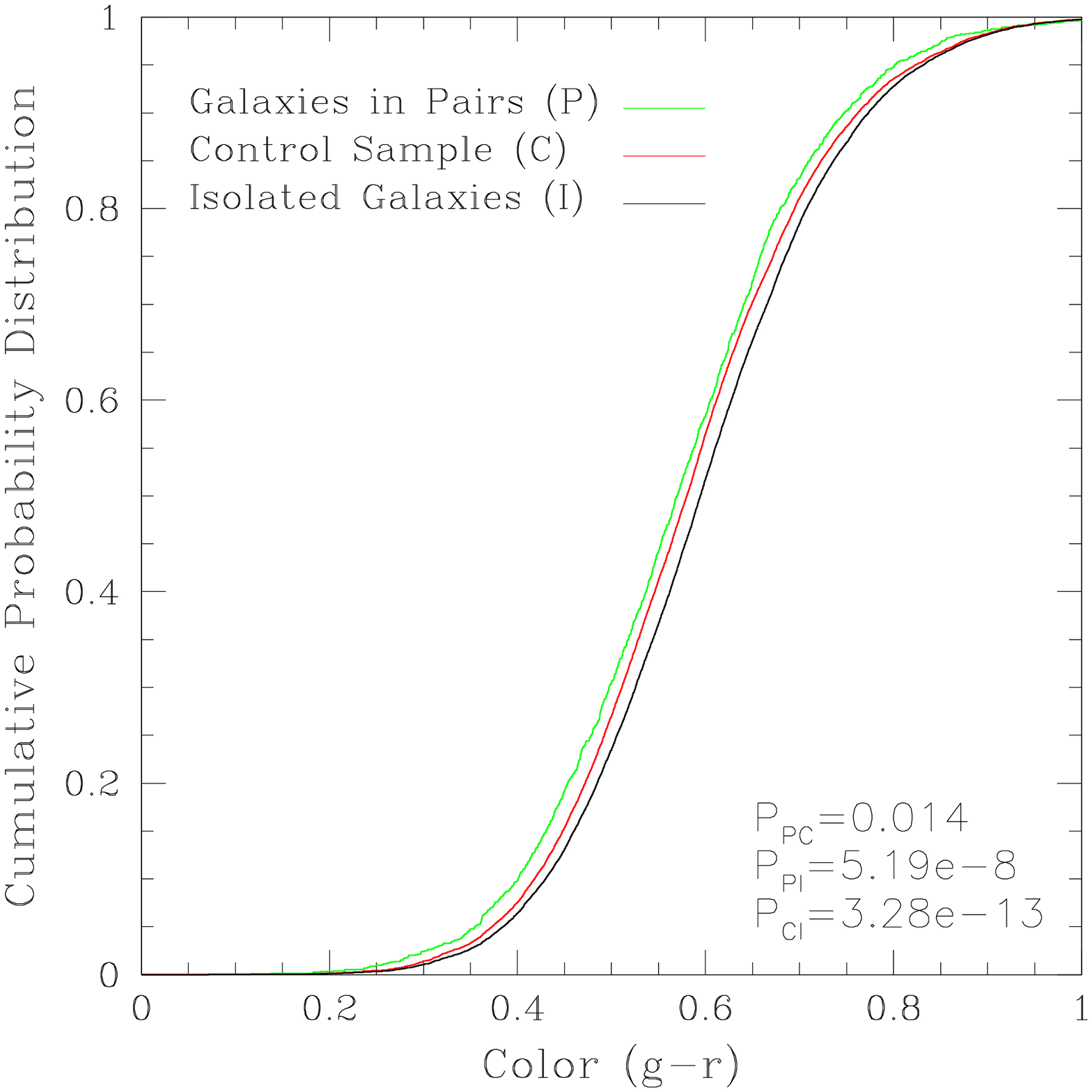}{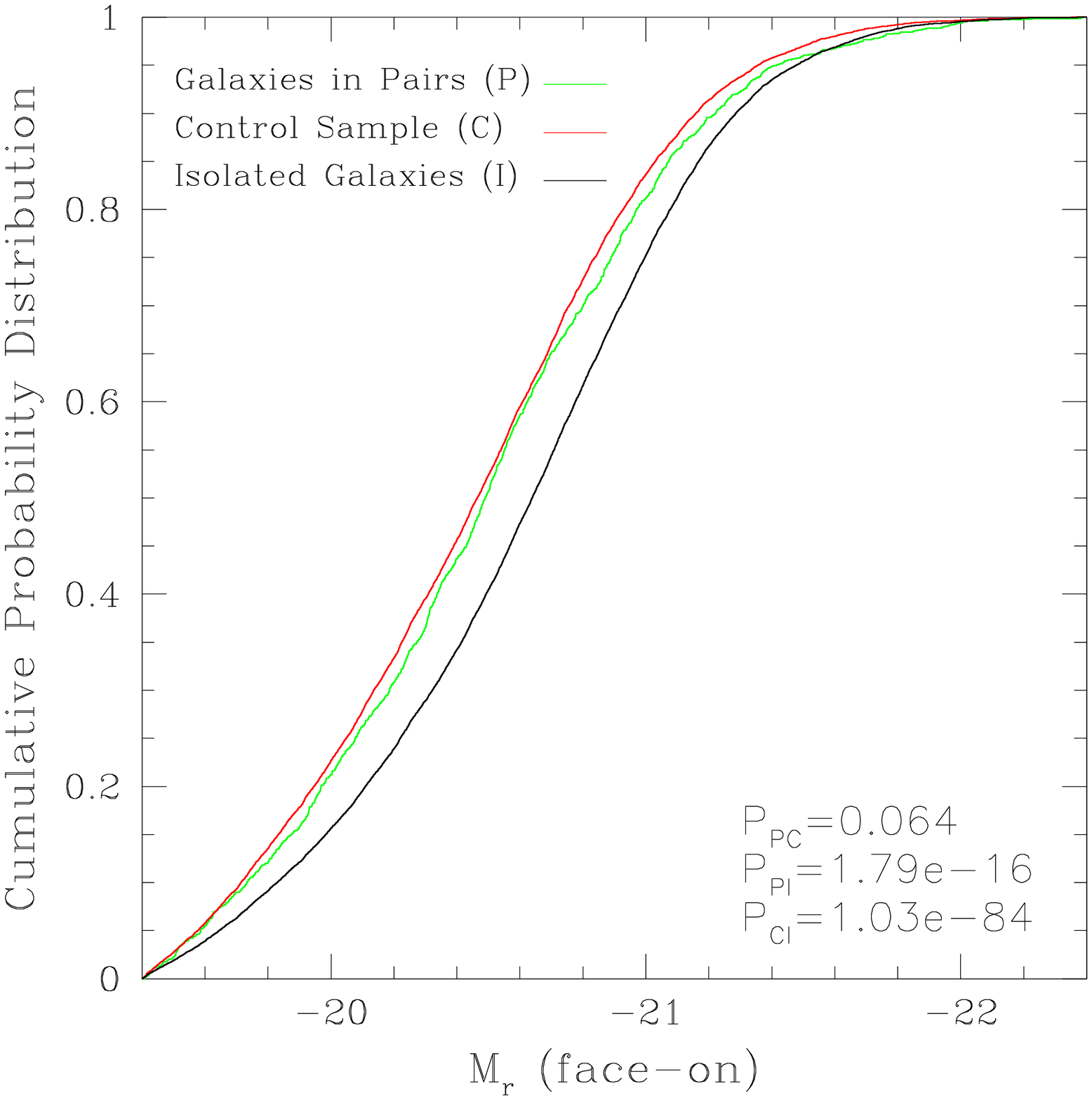}
\caption{Left panel: The cumulative distribution of
$g-r$ color for the $1494$ disk galaxies in isolated pairs
(green), for the matched control sample of $10{,}458$
isolated disk galaxies (red), and for the complete
sample of $16{,}814$ isolated disks (black).
Right panel: The cumulative distribution of
$M_r$, corrected to face-on using the presciption of
\citet{un08} for the disk galaxies in isolated pairs
(green), for the matched control sample of isolated
disks (red), and for all isolated disks (black).
}
\label{fig:2}
\end{figure}

The left panel of Figure~\ref{fig:2} shows
the cumulative distribution of $g-r$ color for
the disks in pairs (green line), for the
matched control sample of isolated disks
(red line), and the complete sample of
isolated disks (black line). The control sample
consists of galaxies that are redder than the
disks in pairs, at a statistically significant
level ($P_{\rm KS} = 0.014$); the median color of
the control sample is $g-r = 0.580$, while the median
color of the disks in pairs is $g-r = 0.570$.
This difference in color may be due in part
to starburst triggered in close encounters;
\citet{pa11} found that ``blue cloud'' galaxies
in close pairs ($R_p < 80 {\rm\,kpc}$, $\Delta v
< 200 {\rm\,km}{\rm\,s}^{-1}$) have $g-r$ colors
that are $\sim 0.02$ magnitudes bluer than a sample
of galaxies, matched to have the same distribution
of redshifts and stellar masses, that are not
in close pairs. Since our sample of pairs contains
primarily wider pairs (out to $R_p = 1 {\rm\,Mpc}$),
it is unsurprising that our color difference, of
$\sim 0.01$ magnitudes, is smaller than that found
by \citet{pa11}.

The right panel of Figure~\ref{fig:2} shows the
cumulative distribution of the corrected (face-on)
absolute magnitude $M_r^f$ for the disks in pairs
(green line), the matched control sample of isolated
disks (red line), and the complete sample of isolated
disks (black line). Although the complete sample of
isolated galaxies is $0.148$ magnitudes brighter, in
the median, than the sample of paired galaxies, the
matched control sample is $0.026$ magnitudes fainter.
For the remainder of this paper, we will compare the
sample of $N_p = 747$ disk galaxies in pairs to the
matched control sample of isolated galaxies, and not
to the complete sample of $16{,}814$ isolated galaxies,
which skew toward higher luminosity, redder color, and
higher redshift.

\begin{figure}
\plotone{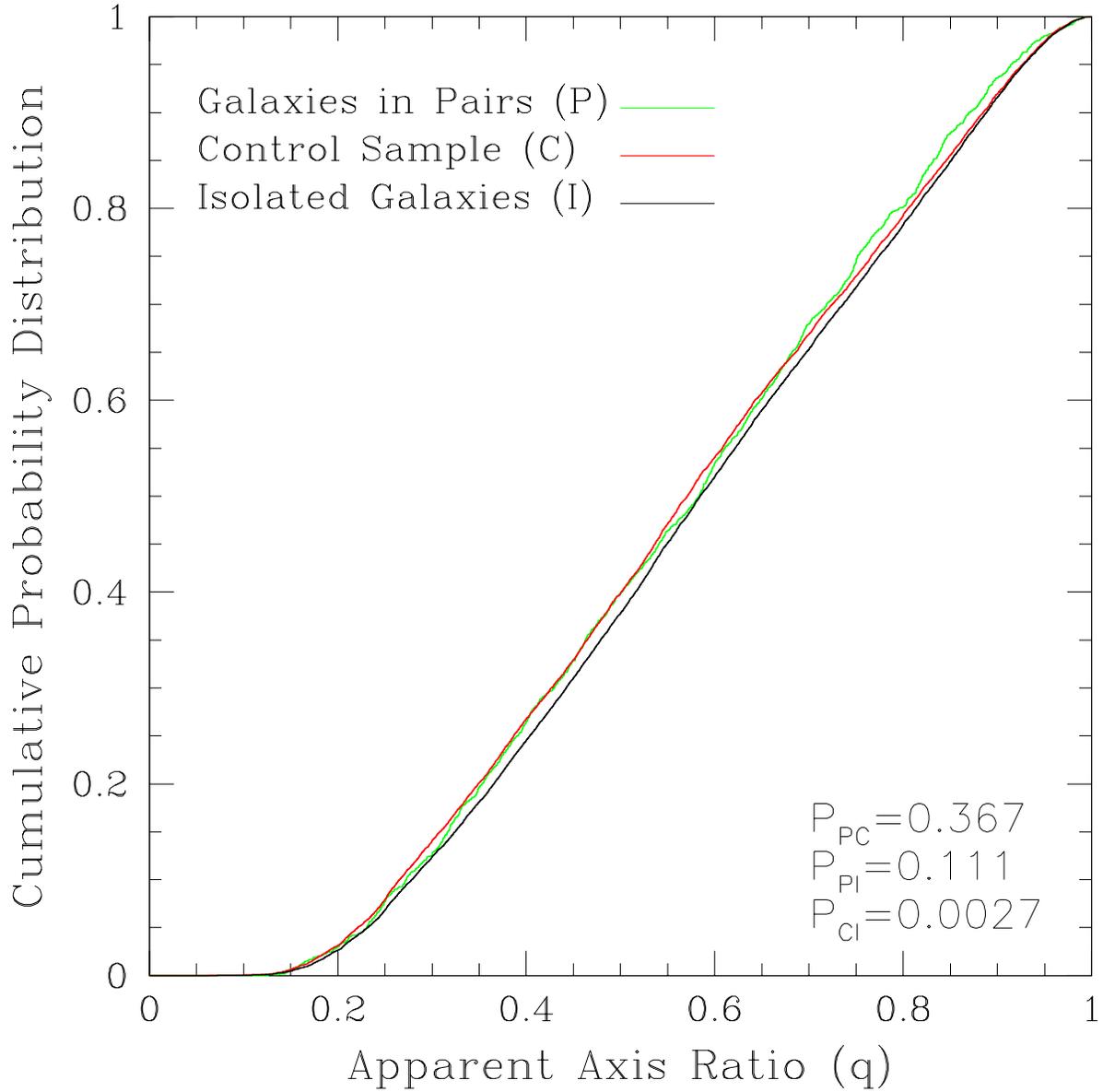}
\caption{The cumulative distribution for the
apparent axis ratios $q$ of the $1494$ disk galaxies
in isolated pairs (green), for the matched control sample of
$10{,}458$ isolated disks (red),  and for the complete sample
of $16{,}814$ isolated disks (black).
}
\label{fig:3}
\end{figure}

The difference in median color between isolated
disk galaxies and disk galaxies in pairs cannot be
attributed to a difference in inclination-dependent
reddening between the two populations.
The green line in Figure~\ref{fig:3} shows the
cumulative distribution of the isophotal axis
ratio $q$ for the $1494$ disk galaxies in pairs. The
red line shows the distribution of $q$ for the
matched control sample of isolated galaxies.
(If spiral galaxies were randomly oriented, infinitesimally
thin, perfectly circular disks, then the distribution would
be a straight diagonal line from lower left to upper
right on the plot; the lack of galaxies at small $q$
indicates they are not infinitesimally thin and the
lack of galaxies at large $q$ indicates they are not
perfectly circular.)
A Kolmogorov-Smirnov test comparing the shapes of
isolated spirals in the matched control sample (red)
and the spirals in binaries (green)
yields a probability score $P_{\rm KS} = 0.367$,
consistent with their being drawn from the same
parent distribution.

\section{ANALYSIS: APPARENT MAJOR AXIS ORIENTATION}
\label{sec-phi}

\begin{figure}
\plotone{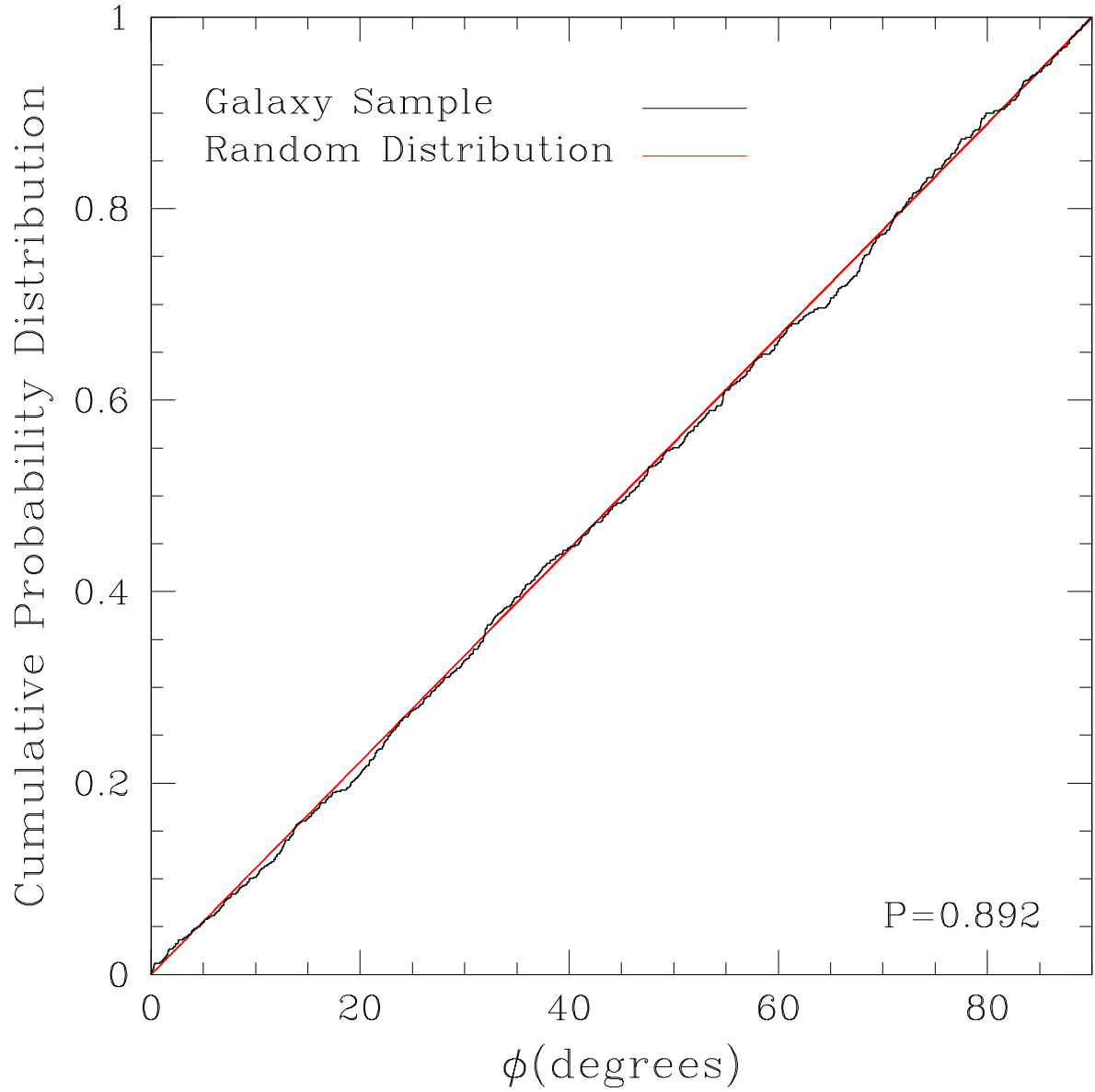}
\caption{The cumulative distribution of the angle
$\phi$ between the major axes of the galaxy images
in the $747$ isolated pairs of disk galaxies in our
main sample (black); for comparison, a uniform distribution
of $\phi$ is shown as the red line. 
}
\label{fig:4}
\end{figure}

The first portion of our analysis involves studying the angle
$\phi$ between the projected major axes of the two galaxy
images in each of our $N_p = 747$ isolated binary spiral systems.
The angle $\phi$ is given by the relation
\begin{equation}
\cos \phi = | \cos ( \theta_1 - \theta_2 ) | \ ,
\end{equation}
where $\theta_1$ and $\theta_2$ are the isophotal position
angle for each galaxy in the pair, taken from the $r$ band
$25 {\rm\,mag}{\rm\,arcsec}^{-2}$ isophote. If the disks of
the two galaxies are randomly oriented with respect to each
other in three dimensions, then the distribution of $\phi$
will be uniform in the interval $[ 0^\circ , 90^\circ ]$
\citep{oo93}. The cumulative probability distribution for
$\phi$ is shown as the black line in Figure~\ref{fig:4};
the red line is the result for a distribution uniform
in the interval $[ 0^\circ , 90^\circ ]$. A Kolmogorov-Smirnov
test comparing the two distributions yields a probability
score $P_{\rm KS} = 0.892$. Thus, the distribution of $\phi$
is consistent with the hypothesis of random disk orientations.

\begin{figure}
\plotone{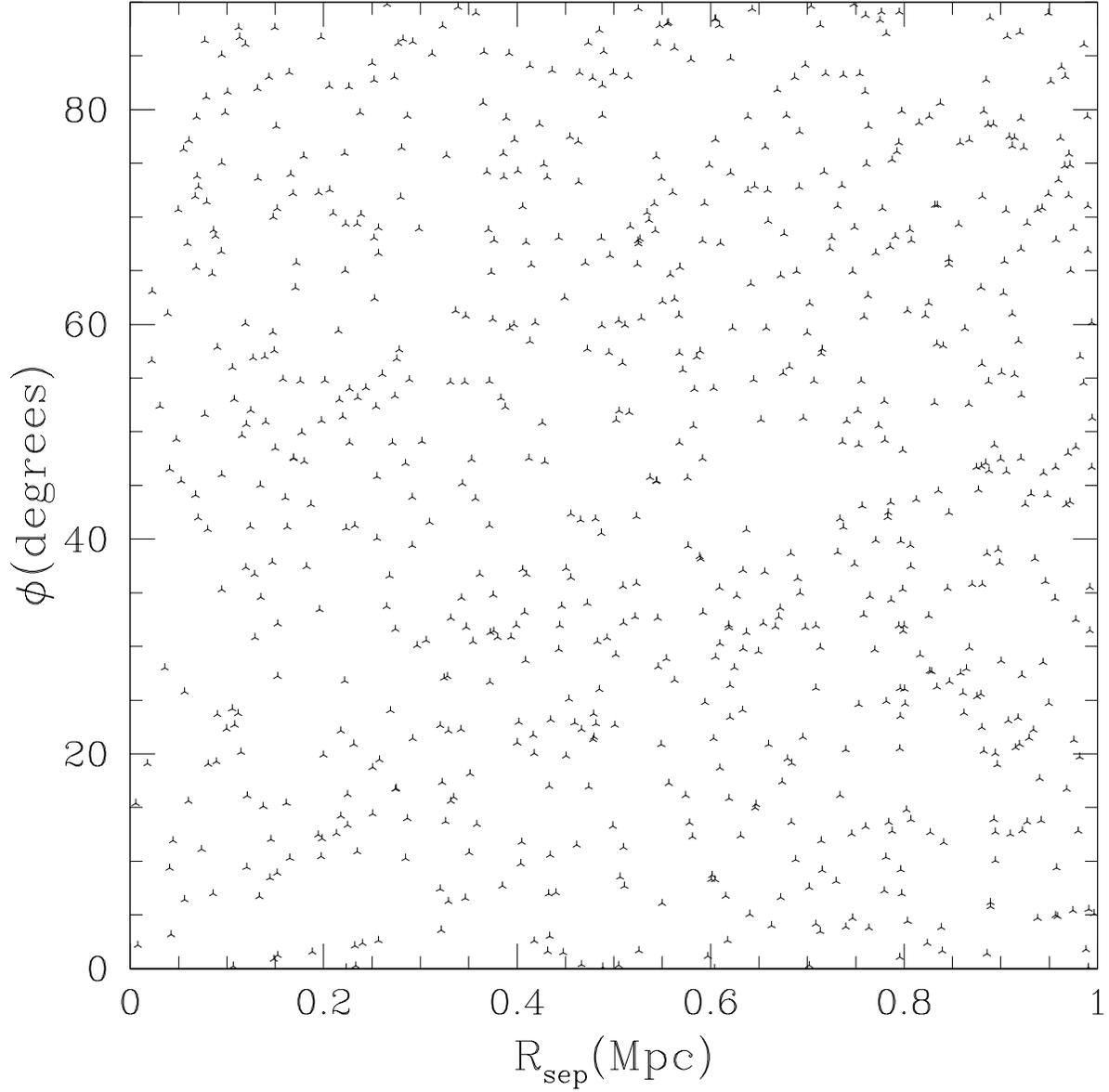}
\caption{The angle $\phi$ between the major axes of the galaxies
images in the $747$ isolated pairs, plotted as a function of
the projected separation $R_p$ between the galaxy centers.
}
\label{fig:5}
\end{figure}

A plot of the angle $\phi$ versus the projected separation
$R_p$ of the galaxies in each isolated binary is presented
in Figure~\ref{fig:5}. The distribution of points reveals
no trends with increasing $R_p$; in particular, close pairs
have no tendency to be more strongly aligned. A KS test
comparing the distribution of $\phi$ for the $N = 335$
pairs with $R_p < 0.5 {\rm\,Mpc}$ to the
distribution for the $N = 412$ pairs with
$R_p \geq 0.5 {\rm\,Mpc}$ yields $P_{\rm KS} = 0.885$,
confirming the visual impression of Figure~\ref{fig:5}.

\section{ANALYSIS: ROTATION AXIS ORIENTATION}
\label{sec-beta}

Although a uniform distribution of $\phi$ is consistent
with random orientations in three dimensions, it doesn't
demand random orientations. In the general case, our
knowledge of the distribution of $\beta$, the angle
between the spin axes of the galaxies, is impeded by
the ambiguity of equation~(\ref{eq:cosbeta}). However,
if we chose only those pairs which have at least one
nearly face-on member, then we can assume an inclination
$\cos i_1 \approx 1$ for one member, and thus
\begin{equation}
\cos \beta \approx \cos i_2 \approx
\left( {q_2^2 - \gamma^2 \over 1 - \gamma^2 } \right)^{1/2} \ ,
\label{eq:face}
\end{equation}
where $\gamma = 0.22$ is the intrinsic short-to-long axis
ratio we use for the second galaxy. If the second galaxy has
$q_2 \leq 0.22$, it is assumed to be edge-on, resulting in
$\cos\beta = 1$.

Similarly, if we chose only those pairs which have at least
one edge-on member, then we can assume $\cos i_1 \approx 0$,
and thus
\begin{equation}
\cos \beta \approx \sin i_2 \cos \phi \approx 
\left( {1 - q_2^2 \over 1 - \gamma^2} \right)^{1/2} 
| \cos ( \theta_1 - \theta_2 ) | \ .
\label{eq:edge}
\end{equation}
Identifying the spiral galaxies that are very nearly
edge-on or face-on, using photometric information alone,
is necessarily approximate, given that spiral galaxies
are not perfectly circular and do not all have the
same intrinsic short-to-long axis ratio. \citet{un08}
fitted the distribution of intrinsic thickness $\gamma$ with
a Gaussian distribution; for luminous ${\rm fracDeV} \leq 0.1$
galaxies in the $r$ band, they found a best fit of
$\gamma = 0.216 \pm 0.067$. Fitting the distribution of intrinsic
disk ellipticity $\varepsilon$ with a lognormal distribution,
\citet{un08} found $\ln \varepsilon = -2.56 \pm 0.91$, implying
a median ellipticity $\varepsilon_{\rm med} = 0.077$.

For the purposes of this study, we assume that a galaxy
with $q \geq 0.9$ is face-on, and a galaxy with $q \leq 0.3$
is edge-on. If spiral galaxies truly had the shape distribution
given by \cite{un08} and quoted above, and were viewed from
random angles, then the galaxies falling into our ``face-on''
class would have $\langle \cos^2 i_1 \rangle = 0.865$ 
and the galaxies in our ``edge-on'' class would have
$\langle \cos^2 i_1 \rangle = 0.019$. Of our 747 isolated
spiral galaxy pairs, $N_f = 94$ pairs contain at least one
face-on spiral and $N_e = 171$ pairs contain at least
one edge-on spiral. Since 14 pairs consist of a face-on
spiral mated with an edge-on spiral, the total number of
pairs with at least one face-on or edge-on spiral is
$N_\beta = 251$.

\begin{figure}
\plotone{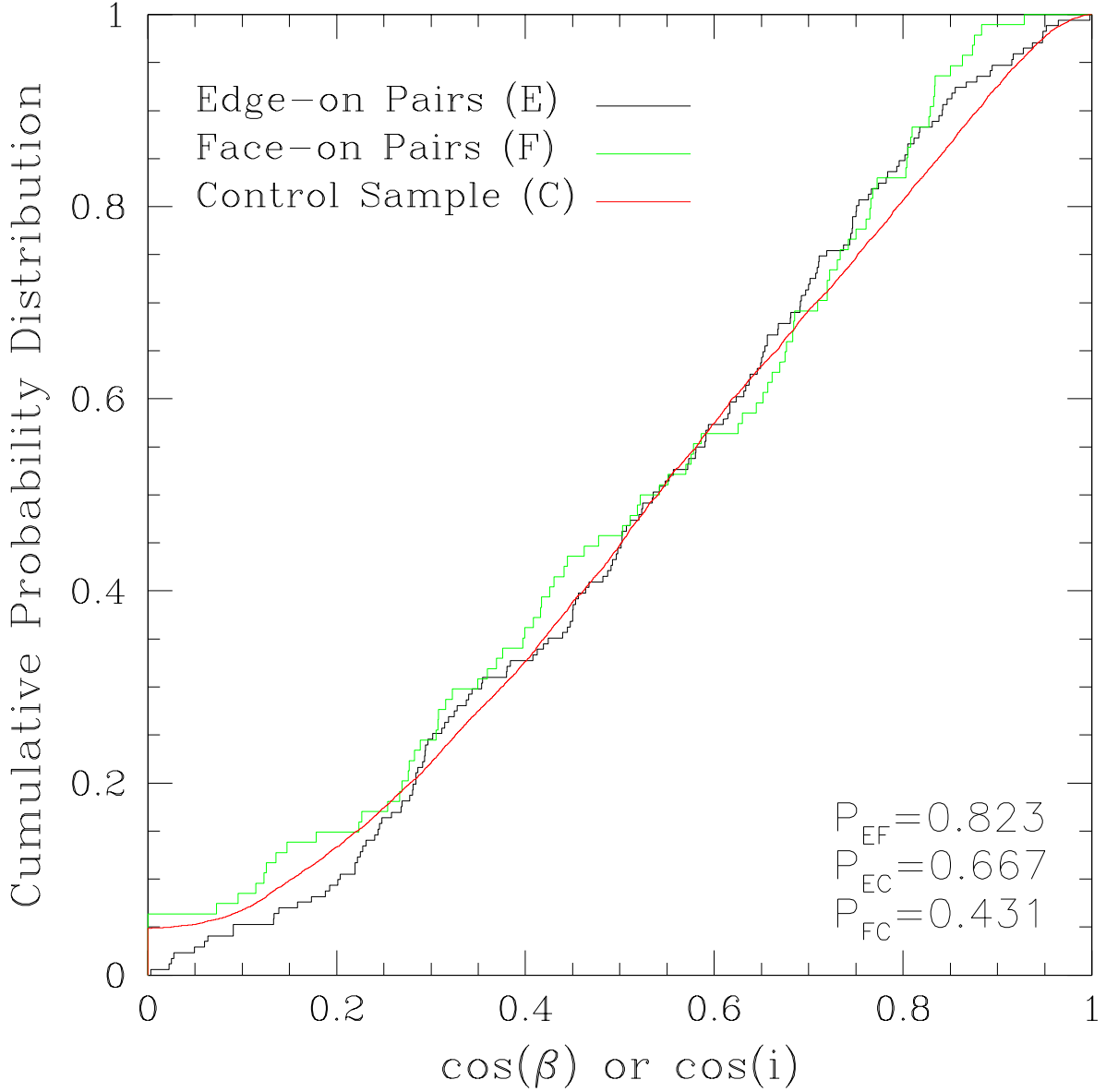}
\caption{The green line shows the cumulative distribution
of $\cos \beta$, estimated from equation~(\ref{eq:face}), for
the $N_F = 94$ isolated pairs in which at least one galaxy
is face-on. The black line shows the distribution of
$\cos \beta$, estimated from equation~(\ref{eq:edge}), for
the $N_E = 171$ pairs in which at least one galaxy is edge-on.
A galaxy is judged to be edge-on if $q \leq 0.3$ and face-on
if $q \geq 0.9$. For comparison, the red line shows the
distribution of $\cos i$, estimated from equation~(\ref{eq:cosi}),
for the matched control sample of isolated disks.
}
\label{fig:6}
\end{figure}

In Figure~\ref{fig:6}, the black line shows the cumulative
distribution function of $\cos\beta$ for the 171 galaxy
pairs containing at least one edge-on spiral; for these
pairs, $\cos\beta$ is computed using equation~(\ref{eq:edge}).
The green line in Figure~\ref{fig:6} shows the cumulative
distribution function of $\cos\beta$ for the 94 pairs
containing at least one face-on spiral; for these pairs,
$\cos\beta$ is computed using equation~(\ref{eq:face}).
Comparing the two distributions with a KS test, we find
$P_{\rm KS} = 0.823$. The two distributions are consistent
with their being drawn from the same parent population of
$\cos\beta$, which is what we expect in the absence of
inclination-dependent selection effects. If the spiral galaxies
in binary systems were randomly oriented with respect to
each other, then the distribution of $\cos\beta$ would
be uniform in the interval $[0,1]$. However, a KS test
reveals that the distributions shown in Figure~\ref{fig:6}
are inconsistent, at a high level of statistical significance,
with their being drawn from a uniform distribution of
$\cos\beta$; for the pairs with an edge-on spiral, $P_{\rm KS}
= 0.021$, and for pairs with a face-on spiral, $P_{\rm KS}
= 0.004$. This is not necessarily an indication that
the spiral galaxies are non-randomly oriented. The approximate
values of $\cos\beta$ computed in equations~(\ref{eq:face})
and (\ref{eq:edge}) assume that spiral galaxies are
perfect oblate spheroids, all with the same intrinsic axis
ratio $\gamma$. These assumptions are known to be inexact.

For a more useful test of the randomness of the orientation
of disks in a pair, we compare the distribution of the
computed values of $\cos\beta$ for the isolated pairs with
the distribution of $\cos i$ for the $N = 10{,}458$ isolated
spiral galaxies in our matched control sample.
The values of $\cos i$, computed using
equation~(\ref{eq:cosi}), contain the same inexact
approximations as the values of $\cos \beta$, computed using
equations~(\ref{eq:face}) and (\ref{eq:edge}). In particular,
the distribution of $\cos i$ for isolated spirals, given
by the red line in Figure~\ref{fig:6}, should be directly
comparable to the distribution of $\cos\beta$ for pairs
including a face-on spiral, given by the green line in
Figure~\ref{fig:6}; in both these cases, we are computing
the cosine of the inclination of a spiral galaxy relative
to the line of sight to that galaxy. However, having a
face-on spiral within $R_p = 1 {\rm\,Mpc}$ doesn't significantly affect
a spiral galaxy's orientation relative to the line of sight;
a KS test comparing the distribution of $\cos i$ for galaxies
in the control sample and $\cos\beta$ for pairs with a face-on galaxy
yields $P_{\rm KS} = 0.431$. 

Our statistical tests are consistent with the hypothesis that the
two disks in an isolated binary spiral system are randomly oriented with
respect to each other. However, since structure on scales larger
than $\sim 1 {\rm\,Mpc}$ has an influence on tidal torques,
we also investigate whether the number density of galaxies on
a larger scale may influence the distribution of $\cos\beta$ for
binaries. For each of our $171$ pairs with a face-on member
and $94$ pairs with an edge-on member, we determine the number of
neighboring galaxies that are within a projected distance
of $R_p = 5 {\rm\,Mpc}$ of either galaxy in the pair and
with a radial velocity difference less than $\Delta v = 300 {\rm\,km}
{\rm\,s}^{-1}$ of either galaxy in the pair. To qualify as a
neighboring galaxy, a galaxy must be in the Sloan spectroscopic
sample, and its $r$ band absolute magnitude cannot be more than
1 magnitude dimmer than that of the lower-luminosity galaxy
in the pair. The median number of neighboring galaxies,
defined in this manner, was six. Consequently, we define pairs
with six or fewer neighboring galaxies to be our ``low-density"
subsample, and pairs with more than six neighboring galaxies
to be our ``high-density" subsample.

\begin{figure}
\plottwo{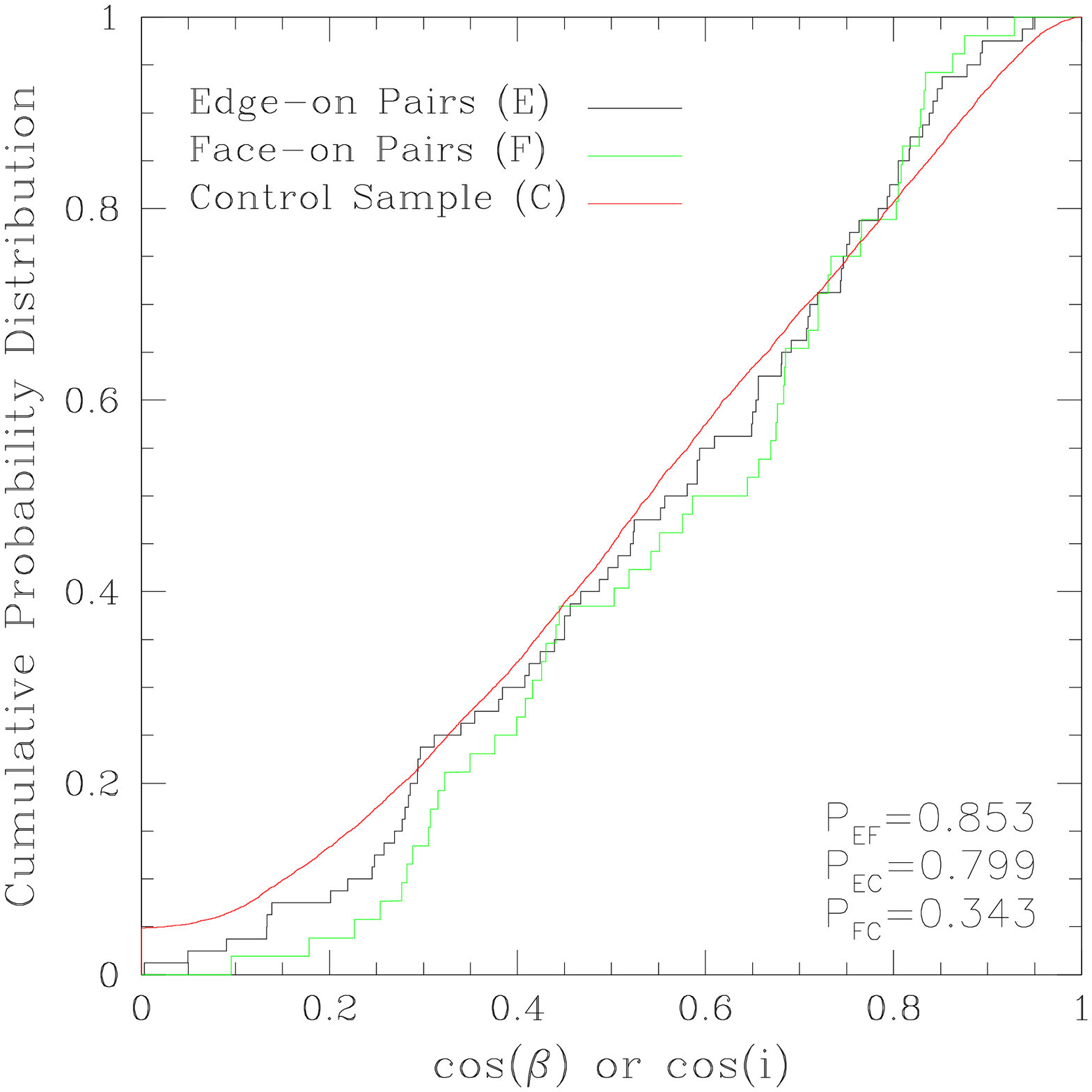}{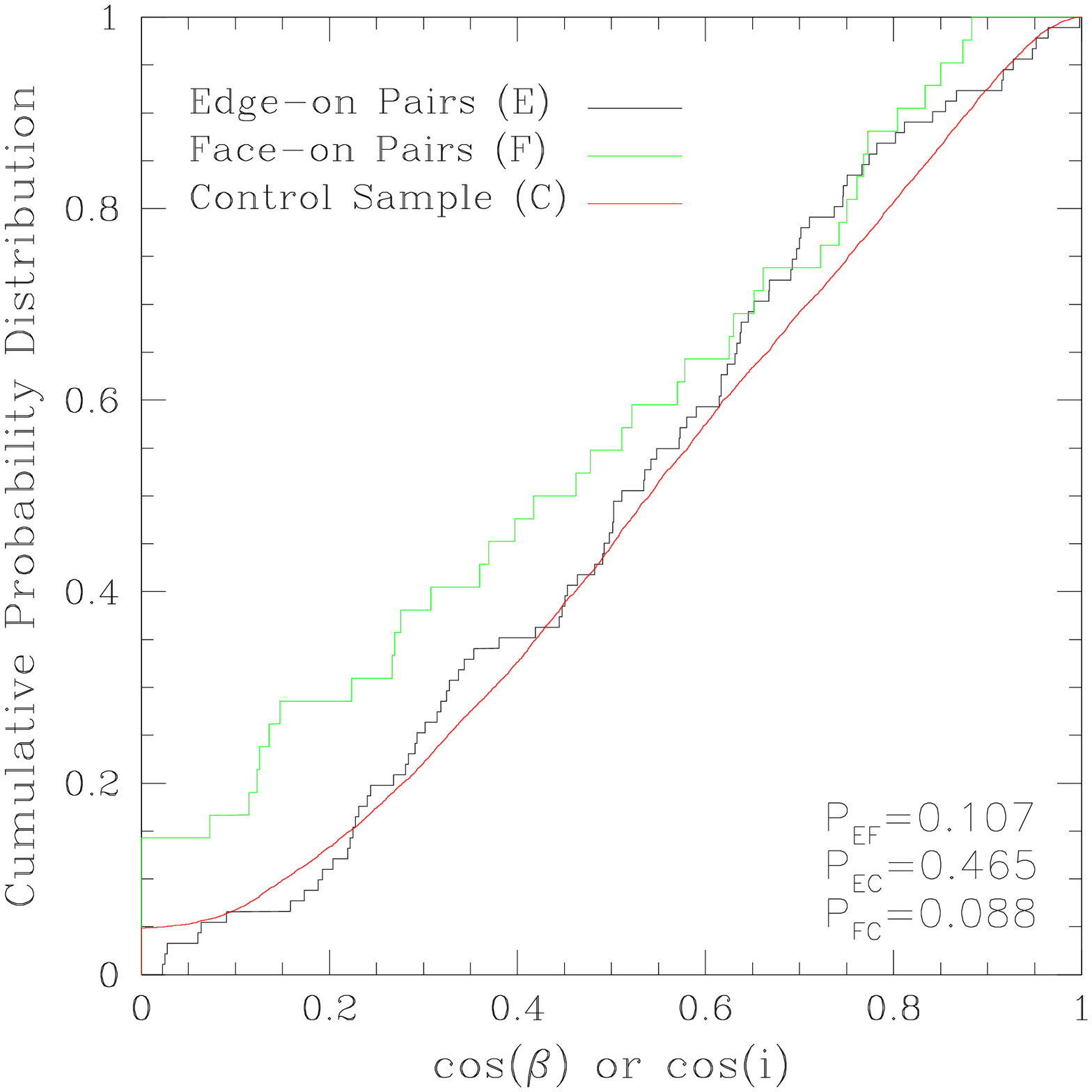}
\caption{The cumulative distribution of $\cos\beta$ for
pairs in the low-density subsample (left panel) and high-density
subsample (right panel). Within each panel, the black line
indicates pairs with at least one edge-on disk and the
green line indicates pairs with at least one face-on disk.
For comparison, the red line shows the distribution of $\cos i$ for
the matched control sample of isolated disks. The values of $\cos\beta$
and $\cos i$ are estimated as described in the caption of Figure~\ref{fig:6}.
}
\label{fig:7}
\end{figure}

The left panel of Figure~\ref{fig:7} shows the distribution of
$\cos\beta$ for pairs in the low-density subsample, while the
right panel of Figure~\ref{fig:7} shows the distribution for the
high-density subsample. In each panel, the distribution of $\cos i$
for the isolated spiral galaxies in the matched control sample
is repeated for reference (red line).
If the simple picture of an isolated binary spiral system having parallel
disks were correct, we might expect the pairs in the low-density
subsample (left panel) to be show higher values of $\cos\beta$.
In fact, we find that the distribution of $\cos\beta$ for both
types of binary spiral systems (ones with edge-on disks and ones with
face-on disks) is indistinguishable from the distribution of
$\cos i$ for isolated spirals. The only marginally significant result we find,
at the $\sim 90\%$ confidence level, is that in high-density regions
(right panel of Figure~\ref{fig:7}) binary spiral systems containing
a face-on disk tend to have a distribution of $\cos\beta$ that is different
from the distribution of $\cos i$ for isolated spirals: $P_{\rm KS} = 0.088$.
The difference is such that the partners of face-on disks are more likely
to be seen edge-on than an isolated disk would be. In fact, of the 42 partners
of face-on disks located in high-density regions, 6 have $q_2 < 0.22$.
Of the isolated disks in the control sample,
4.9\% have $q < 0.22$; given a probability $P=0.049$
of an individual disk being edge-on, the expected probability
of finding 6 or more edge-on galaxies in a sample of 42 would
be $P = 0.016$, using the standard binomial formula. Although,
in high-density regions, the partners of face-on galaxies have
a high probability of being edge-on, the partners of edge-on
galaxies do not have a comparably high probability of being face-on.

\section{DISCUSSION}
\label{sec-disc}

In summary, our study of disk galaxies in the
Sloan Digital Sky Survey reveals no strong evidence for
a preferred alignment of the individual disk galaxies
in a binary galaxy system. For 747 close pairs of galaxies,
the distribution of $\phi$, the angle between the galaxies'
projected major axes, is consistent with a uniform distribution
between $0^\circ$ and $90^\circ$. This is consistent with
disks that are randomly oriented with respect to each other.

Looking solely at $\phi$, however, doesn't use all the
available information. Computing $\cos\beta$, where $\beta$
is the angle between the rotation axes of the disks, makes
use of additional information -- the apparent axis ratios
of the two galaxies
(equations~\ref{eq:cosi} and \ref{eq:cosbeta}).
Taking advantage of the loss of tilt ambiguity when
one disk is face-on or edge-on, we can compute the
distribution of $\cos\beta$ for these special pairs.
Looking at 171 close pairs with an edge-on galaxy
and 94 close pairs containing a face-on galaxy reveals
no statistically significant tendency for the two
disk galaxies in a pair to be either parallel to
each other ($\cos\beta \sim 1$) or perpendicular
to each other ($\cos\beta \sim 0$). 

We find one curious result that is consistent with
correlated orientations. In relatively high-density
neighborhoods (more than 6 bright neighboring galaxies
with $R_p \leq 5 {\rm\,Mpc}$), face-on galaxies within a
close pair have an unusually high probability of having
a edge-on partner; although just 4.9\% of all disk galaxies
in our sample have $q < 0.22$, 6 out of 42 partners of
face-on galaxies in high-density neighborhoods have $q < 0.22$.
The physical significance of this statistically significant
result is unclear.

Obviously, the toy model for angular momentum acquisition,
in which a pair of protogalaxies spin each other up and
have antiparallel spin vectors and a tiny orbital angular
momentum, is inconsistent with our results. Acquisition of
spin angular momentum is supplied by torques from an array
of protogalaxies and protogroups in the vicinity. Moreover,
the initial angular momentum of a disk, acquired by tidal
torques, can be modified by encounters and late infall.
The continuing process of acquiring and modifying angular
momentum appears, from our results, to produce binary
spiral galaxies that have random relative orientations of
their stellar disks.


We thank the referee for valuable advice.
Funding for the SDSS and SDSS-II has been provided
by the Alfred P. Sloan Foundation, the Participating Institutions,
the National Science Foundation,
the U.S. Department of Energy, the National Aeronautics
and Space Administration, the Japanese Monbukagakusho,
the Max Planck Society, and the Higher Education Funding Council
for England. The SDSS website is \url{http://www.sdss.org/}.
The SDSS is managed by the Astrophysical Research Consortium (ARC) for
the Participating Institutions. 
The Participating Institutions are the American Museum of Natural History,
Astrophysical Institute Potsdam, University of Basel,
University of Cambridge, Case Western Reserve University,
University of Chicago, Drexel University, Fermilab,
the Institute for Advanced Study, the Japan Participation Group,
Johns Hopkins University, the Joint Institute for Nuclear Astrophysics,
the Kavli Institute for Particle Astrophysics and Cosmology,
the Korean Scientist Group, the Chinese Academy of Sciences (LAMOST),
Los Alamos National Laboratory, the Max-Planck-Institute for
Astronomy (MPIA), the Max-Planck-Institute for Astrophysics (MPA),
New Mexico State University, The Ohio State University,
University of Pittsburgh, University of Portsmouth, Princeton University,
the United States Naval Observatory, and the University of Washington.

\end{document}